\documentclass[aps,prl,showpacs,nobibnotes,twocolumn,preprintnumbers,
nobalancelastpage,superscriptaddress]{revtex4}
\usepackage{latexsym}
\usepackage{dcolumn}
\usepackage{bm}
\input{epsf}
\newcommand{\be}{\begin{equation}}
\newcommand{\ee}{\end{equation}}
\newcommand{\ba}{\begin{eqnarray}}
\newcommand{\ea}{\end{eqnarray}}
\newcommand{\bfi}{\begin{figure}
\epsfxsize=9cm
\epsffile}
\newcommand{\efi}{\end{figure}}
\newcommand{\la}{\lesssim}
\newcommand{\ga}{\gtrsim}

\begin{document}

\title{Testing Gravity Against Early Time Integrated Sachs-Wolfe Effect}
\author{Pengjie Zhang}
\email{pjzhang@shao.ac.cn}
\affiliation{Shanghai Astronomical Observatory, Chinese Academy of
  Science, 80 Nandan Road, Shanghai, China, 200030}
\affiliation{Particle Astrophysics Center,
Fermi National Accelerator Laboratory, Batavia, IL 60510-0500, USA}
\begin{abstract}
A generic prediction of general relativity is that the cosmological
linear density growth factor $D$ is scale independent.  But in general,
modified gravities do not preserve this signature. A scale dependent
$D$ can cause time variation in gravitational potential at high
redshifts and provides a new cosmological test of gravity, through
early time integrated Sachs-Wolfe (ISW) effect-large scale 
structure (LSS)  cross correlation.  We demonstrate the power of this
test for a class of  $f(R)$ gravity, with the form  $f(R)=-\lambda_1
H_0^2\exp(-R/\lambda_2H_0^2)$.  Such $f(R)$ gravity, even with
degenerate expansion history to $\Lambda$CDM,  can produce
detectable ISW effect at $z\ga 3$ and $l\ga 20$. Null-detection of
such effect would constrain $\lambda_2$ to be $\lambda_2>1000$ at
$>95\%$ confidence level.  On the
other hand, robust detection
of  ISW-LSS cross correlation at high $z$ will severely challenge general 
relativity. 
\end{abstract}
\ \maketitle

{\it Introduction}.---
Cosmological observations provide unique tools to study gravity at
$\ga$ Mpc scales.  General relativity, with the aid of the
cosmological constant, or dark energy with  equation of state $w\sim
-1$, successfully reproduces the accelerated expansion of the
Universe, indicated by SN Ia 
observations\cite{SNIa}, along with the flatness of the Universe
measured by the cosmic microwave background (CMB)\cite{WMAP} and
distance measured by the baryon oscillations\cite{Eisenstein05}.
However, these 
observational evidences mainly constrain the mean expansion history of
the Universe and can be 
reproduced  by modified gravity such as brane world DGP theory
\cite{DGP} and 
generalized $f(R)$ gravity\cite{Carroll1}.  Essentially, the large
scale structure (LSS) of the universe,  such as weak
gravitational lensing, galaxy clustering and the
integrated Sachs-Wolfe (ISW) effect\cite{White01,Lue04a,Lue04b,Sealfon05,Shirata05,Knox05},
is required to break this 
degeneracy.

General relativity imprints a unique signature in the LSS, which is
scale {\it independent} linear density  growth factor $D$  at
sub-horizon   
scale after matter-radiation equality
epoch\cite{Note_Neutrino}. Modifications to general relativity not
only changes  the amplitude of $D$, but in general,  causes $D$ to be scale  
dependent. This unique feature  of modified gravity has already been
noticed in  phenomenological theory of 
modified Newtonian 
potential\cite{White01,Shirata05,Sealfon05}. It can be
detected by weak gravitational lensing\cite{White01}, galaxy clustering
\cite{Shirata05,Sealfon05} and late time ISW
effect. Counter-intuitively, in this paper, we show that
modified gravity  can  produce a detectable {\it early time} ISW effect.

We investigate a class of $f(R)$ gravity with action
\be 
L=\int (R+f(R))\sqrt{g}d^4x+L_{\rm matter}\ ,
\ee
and field equation
\be
\label{eqn:field}
(1+f_R)R_{uv}-\frac{g_{uv}}{2}(R+f-2\Box f_R)-f_{R;u;v}=8\pi
GT_{uv}\ ,
\ee
where $f_R\equiv df/dR$.  We design $f(R)=-\lambda_1 H_0^2\exp(-R/\lambda_2H_0^2)$, where  $\lambda_{1,2}$   
are two positive dimensionless constants and $H_0$ is the Hubble constant at
present.   To mimic a $\Lambda$CDM universe, $\lambda_1\sim 1$ is
required. To reduce to the general relativity in the solar system and
pass the solar  system test, $f\ll R$ is required. In this
limit, we have $R\rightarrow 8\pi G \rho_{\rm solar}$, where
$\rho_{\rm solar}$ is the local density where solar system tests are
carried out. In this limit, $f(R)/R \sim [\rho_c/\rho_{\rm
    solar}]\exp[-3\rho_{\rm 
  solar}/\lambda_2\rho_c]$. Since $3\rho_{\rm solar}/\rho_c\ga 
10^{6}$ \cite{Solar}($\rho_c$ is the critical density of the
Universe), models with $\lambda_2\ll 10^6$ can survive all solar system
tests.  For example, For $\lambda_2=10^3$,
  this correction is of the order $\sim
  10^{-400}$. Given such tiny $f(R)$, we expect that $f$, $f_R$,
  $\Box f_R$ and $f_{R;\mu;\nu}$ in Eq. \ref{eqn:field} can all  be
  safely neglected for any physical purpose.

For the $f(R)$
gravity,  the application of Birkhoff 
theorem to perturbations of a spherically symmetric region leads to
scale independent $D$ \cite{Amarzguioui05}.  
We reinvestigate this issue  by solving the 
structure  evolution of the fully covariant $f(R)$ gravity to linear
order in the metric perturbation. We  find that $D$ shows nontrivial scale
dependence, consistent with  the results based on the
Palatini approach \cite{Koivisto05}.

{\it The $H$-$z$ relation of the $f(R)$ gravity}.---
Cosmological observations prohibit strong
deviation of $f(R)$ from the cosmological constant. At the limit
that $R(z=0)\ll \lambda_2 H_0^2$, the $H$-$z$ relation of $f(R)$
gravity can have the same asymptotic behavior as that of
$\Lambda$CDM. At low redshift where $R(a) \ll \lambda_2 H_0^2$, $f(R)$
behaves as a cosmological constant and  the $H$-$z$ relation resembles
that of $\Lambda$CDM. At
high redshifts where $R\gg \lambda_2 H_0^2$, $f(R)\rightarrow 0$ and
$H(z)\rightarrow \Omega_0^{1/2}(1+z)^{3/2}$. Deviation from
$\Lambda$CDM happens at some  intermediate redshifts where
$R(a)\sim \lambda_2 H_0^2$ and vanishes toward both higher and lower $z$.
We quantify their difference by
solving Eq. \ref{eqn:field} of a flat universe to zero order
\be
\label{eqn:expansion}
H^2+\frac{f}{6}-\frac{\ddot{a}}{a}f_R+H\dot{f}_R=H_0^2\Omega_0 a^{-3} \ .
\ee
Here, $a\equiv 1/(1+z)$ is the scale factor. 
This equation can be rewritten as $y=\Omega_0-C(y(a))$, where
$y\equiv a^3H^2$, $C(y(a))\equiv
[f/6-\ddot{a}f_R/a+H\dot{f}_R]a^3$ and $\Omega_0$
is the dimensionless matter density at
present. Since $C(y(a))$ is completely determined once $y$ as a
function of $a$ is given, Eq. \ref{eqn:expansion} can be solved
iteratively by the 
iteration relation
$y^{(i+1)}=\Omega_0-C(y^{(i)})$.  To mimic a $\Lambda$CDM universe, we
fix $\lambda_1$ by requiring $f(R(a=1))=-6H_0^2(1-\Omega_0)$. The
iteration converges quickly by 
taking the initial guess $y^{(0)}=\Omega_0+(1-\Omega_0)a^3$. For
$\lambda_2\geq 100$, $y^{(1)}$ is accurate to $\sim 1\%$. 
As expected, for  $\lambda_2\geq 100$,  the $H(z)$-$z$ relation is almost
identical to the corresponding $\Lambda$CDM cosmology
(Fig. \ref{fig:H}).  Such $f(R)$ gravity can not be distinguished from
$\Lambda$CDM by inflation, big bang nucleosynthesis (BBN), primary
CMB, SN Ias and other measures of $H$-$z$ relation.  

\begin{figure}
\epsfxsize=9.5cm
\epsffile{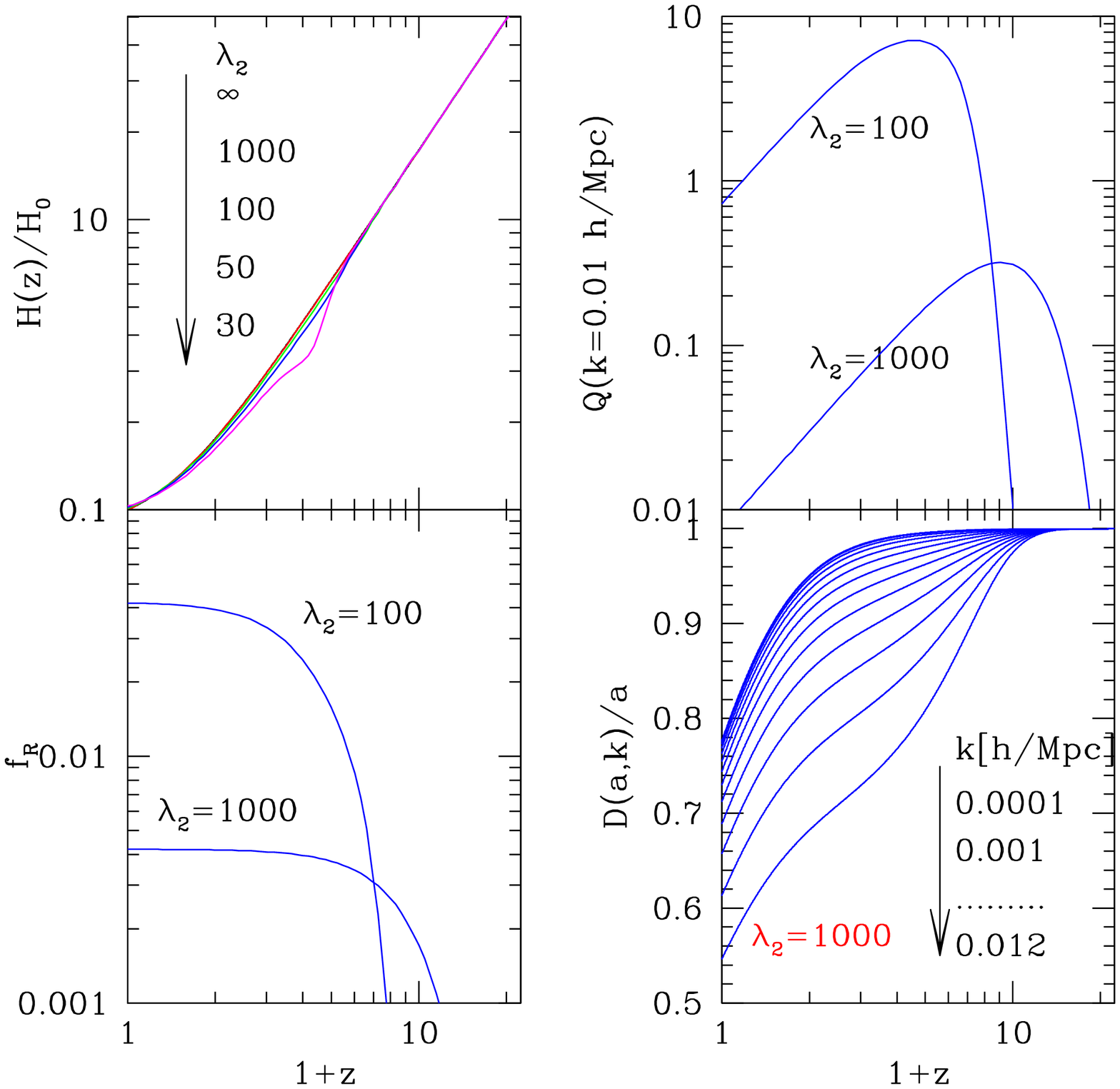}
\caption{The $H(z)$-$z$ relation and structure growth in the exponential
  $f(R)$ gravity. Top left panel: $H$-$z$. $\lambda_2\rightarrow \infty$
  corresponds to $\Lambda$CDM cosmology.
Top right panel: $Q(k,a)\propto k^2$, which describes the main effect
  of $f(R)$ gravity to structure formation.  We plot the result of
  $k=0.01h/$Mpc. Bottom left panel:
  $f_R(a)$, which determines the effective Newton's 
  constant $G_{\rm eff}=G/(1+f_R)$.  For $\lambda_2\ga 100$, its
  effect to structure formation can be neglected.
 Bottom right panel: $D(k,a)/a$
  ($\lambda_2=1000$), where the linear density growth factor $D$ is
  normalized such that $D\rightarrow 
  a$ when $a\rightarrow 0$.
\label{fig:H}}
\efi

{\it The large scale structure of the $f(R)$ gravity}.--- We will show
that, even with
this degeneracy in $H$-$z$  
relation and solar system behavior, the LSS of the $f(R)$ gravity
could be significantly different to that of $\Lambda$CDM. We choose the Newtonian gauge
\ba
ds^2=-(1+2\psi)dt^2+a^2(1+2\phi)\sum_{i=1}^3 (dx^{i})^2 \ .
\ea
There are four perturbation variables $\phi$, $\psi$, the matter
over-density $\delta$ and the (comoving) peculiar velocity 
convergence $\theta$.
 
In general relativity, $\phi=-\psi$, as long as there is no anisotropic
stress. But in modified gravity, this relation breaks in
general.  $ij$ ($i\neq j$) component of  Eq. \ref{eqn:field} provides the
 relation   between $\phi$ and $\psi$. For $f(R)$ gravity, due to 
non-vanishing $f_{R;i;j}$  ($i\neq j$), $\phi$-$\psi$ relation 
becomes scale dependent.   Throughout this paper, we neglect  time
derivative terms with respect to spatial derivative terms of corresponding
variables. This 
simplification holds at scales $k\ga aH \la  
10^{-3}h/$Mpc. Since we will focus on the ISW effect at $l\ga 20$ and
$z\ga 3$ where the relevant $k\ga 5\times 10^{-3}h/$Mpc,  this
simplification is sufficiently accurate. We then obtain 
\be 
\label{eqn:phi-psi}
\phi+\psi=\frac{f_{RR}c^2}{1+f_R}\frac{2}{a^2}(\nabla^2\psi+2\nabla^2\phi)
\ .
\ee
  In Fourier space, this reads $\psi=-\phi(1-2Q)/(1-Q)$, where $Q(k,a)\equiv
-2f_{RR}c^2k^2/(1+f_R)a^2$ and $f_{RR}\equiv d^2f/dR^2$.  For clarity,
we explicitly show the speed 
of light $c$. We will see that this scale dependent
$\psi$-$\phi$ relation has profound effect on the LSS.  Combining
Eq. \ref{eqn:phi-psi} and the $tt$ 
component of  Eq. \ref{eqn:field}, we
obtain the new Poisson equation
\be
\nabla^2(\phi-\psi)=-\frac{3H_0^2\Omega_0}{1+f_R}a^{-1}\delta\ .
\ee
The energy-momentum tensor is still conserved and provides the
remaining two equations:
\ba
\dot{\delta}+\theta=0\ ,\ \ \dot{\theta}+2H\theta+\frac{1}{a^2}\nabla^2\psi=0\ .
\ea
Combining all 4 equations, we obtain the main equation of this paper:
\be
\label{eqn:D}
\delta^{''}+\delta^{'}(\frac{3}{a}+\frac{H^{'}}{H})-\frac{\delta}{a^2}
\frac{1-2Q}{2-3Q}\frac{3H_0^2\Omega_0}{a^3H^2(1+f_R)}=0\ ,
\ee
where $'\equiv d/da$. In general relativity, $Q=0$, so the linear
density growth factor $D\propto \delta(a)/\delta(a_i)$ is scale
independent at scales $k\ga aH/c$, no matter what the form of dark
energy is. Here, $a_i$ is the scale factor at some early epoch and we
normalze $D$ such that $D\rightarrow a$ when $a\rightarrow 0$. But in $f(R)$,
the scale dependent $Q(k,a)$ induces  nontrivial scale dependence to
$D$. 
This behavior  can not be
obtained by  a simple change in  the effective Newton's
constant. Furthermore, the correction $Q$  has a nontrivial
dependence on $a$. This is hard to realize by simply changing the form of the
Newtonian potential (e.g. to Yukawa potential). 

 Since $f_{RR}<0$, there exist one {\it apparent  singularity} $Q=2/3$
 in Eq. \ref{eqn:D}, where only $\delta=0$ solution is accepted and
 two  at $Q=1/2,1$ in the $\psi$-$\phi$ 
 relation, where only $\psi=\phi=0$ solution is accepted. We leave 
 this issue alone 
 until the discussion section. For the moment, we take a modest goal
 by only using 
 regions where $Q< 1/2$ to constrain $f(R)$. For
 $\lambda_2=1000$, this constrains us to region where $k\leq 0.012
 h/$Mpc.  

Hereafter, we fix $\lambda_2=1000$. At $z\gg 1$, $H\propto a^{-3/2}$, $D\propto
a^{1-\eta}$ when $\eta\equiv 3Q/5(2-3Q)\ll 1$. Thus gravitational potential  decays at high
 redshifts with rate $\propto a^{-\eta}$ and causes an observable
 integrated Sachs-Wolfe (ISW) effect. At later time when $R\la
 \lambda_2H_0^2$, $Q\rightarrow 0$ (Fig. \ref{fig:H}), the evolution
 of $D$ approaches that of $\Lambda$CDM.  For the exponential $f(R)$,
 $Q(a)$ peaks at 
 $z\gg 1$ (Fig. \ref{fig:H}), so the resulting ISW effect peaks at $z\gg
 1$, as contrast to that of $\Lambda$CDM cosmology or dark energy
 models with $w\sim -1$. This provides us a unique way to test this form
of $f(R)$.  We solve Eq. \ref{eqn:D} 
 numerically. Initial condition is set to normalize $D\rightarrow a$
 when $a\rightarrow 0$. 

 \bfi{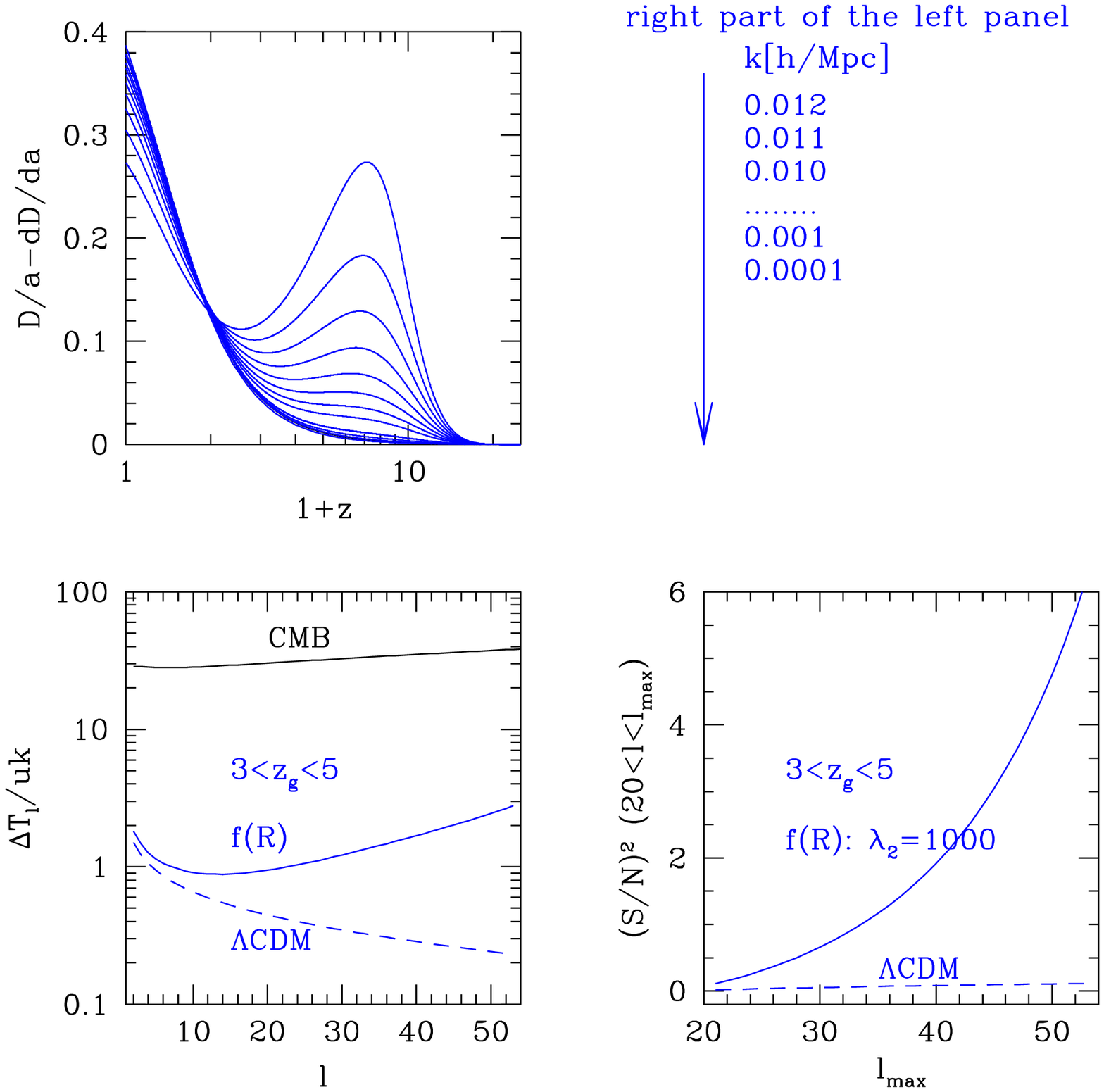}
\caption{The ISW effect. $\lambda_2=1000$ is adopted. Top left panel:
  $D/a-dD/da$, which determines the sign and amplitude of the ISW 
  effect. $D$ is normalized such that $D\rightarrow a$ when
  $a\rightarrow 0$.  Bottom left
  panel: the ISW effect. Bottom right panel: Cumulative  S/N of the
  ISW-LSS cross correlation measurements.  \label{fig:ISW}} 
\efi

{\it The integrated Sachs-Wolfe  effect}. ---
Time variation in $\psi-\phi$ causes a fractional CMB temperature variation \cite{ISW}
\be
\frac{\Delta T}{T_{\rm CMB}}=\int [\dot{\psi}-\dot{\phi}]ad\chi \ .
\ee
Here, $\chi$ is the comoving angular diameter distance. 
Since both  $\psi-\phi$ and the LSS trace the underlying matter
distribution, there exists an ISW-LSS cross correlation, with power spectrum 
\be
\frac{l^2}{2\pi}C_l^{\rm ISW-LSS}=\frac{\pi}{l}\int
\Delta^2_{(\dot{\psi}-\dot{\phi})\delta_{\rm LSS}}(\frac{l}{\chi}) W_{\rm
  LSS}(\chi)a^2\chi d\chi \ . 
\ee
Here, $\delta_{\rm LSS}$ is the density fluctuation of the  LSS
tracers, $W_{\rm LSS}$ is the 
corresponding weighting function and
$\Delta^2_{(\dot{\psi}-\dot{\phi})\delta_{\rm LSS}}$ is the
corresponding 3D power spectrum(variance). The above formula adopts
the Limber's approximation, which is  sufficiently accurate to serve
for our interest  at $l\geq 20$. 
The  amplitude and sign of the ISW effect is determined by $A_{\rm ISW}\equiv 
D/a-dD/da$. Positive $A_{\rm ISW}$ means positive correlation between
ISW and LSS. For $k\ga 0.007h/$Mpc, $A_{\rm ISW}$ has a bump at $z\sim
6$, whose amplitude increases towards small scales (large $k$). This
boosts early time  small scale ISW signal (Fig. \ref{fig:ISW}). 

The S/N of the ISW-LSS cross correlation measurement of each $l$ mode is 
\ba
\left(\frac{S}{N}\right)^2=\frac{(2l+1)f_{\rm sky}C^2_{\rm ISW-LSS}}{(C_{\rm
      CMB}+C_{\rm ISW}+C_{\rm
    CMB}^{\rm shot})(C_g+C_g^{\rm shot})+C^2_{\rm ISW-LSS}}
\ea
Here, $C_{\rm CMB}, C_{\rm ISW},C_g$ are the power spectra of  primary
CMB, ISW, and galaxies, respectively, while $C_{\rm CMB}^{\rm shot}$ and $C_g^{\rm
  shot}$ are the power spectra of associated shot noises,
respectively. Since the exponential $f(R)$ does not affect  physics at 
$z\ga 100$, we adopt the same primordial  power spectrum with power
index $n=1$, the same transfer function BBKS \cite{BBKS} and the same amplitude at
$a_i=0.01$, as that of the $\Lambda$CDM cosmology. 
 The LSS tracers we choose are 21cm emitting galaxies at $3<z<5$,
 which will be measured by proposed 
21cm experiments such as Square Kilometer Array\cite{SKA}. 
 Singularities presented in the
perturbation equations limit us to $l< 60$, where one can neglect
shot noises of CMB. For the estimations of  LSS clustering signal and shot noise,
biggest uncertainties are (1) HI ({\it neutral hydrogen}) mass function at
$3<z<5$, (2) 21cm emitting
galaxy bias and (3) specifications of 21cm experiments.  If one adopts HI mass
functions calibrated against observations of damped
Lyman-$\alpha$ systems and Lyman limit systems, SKA can detect $\ga  10^9$ galaxies
at $z>3$ in five years across the whole sky, for a field of view $\ga 10$ deg$^2$ at $\sim 300$ Mhz (for
details of the calculation, see, e.g. \cite{Zhang05}). Detection
thresholds of  HI mass at $z\ga 3$ are $\ga 10^9M_{\odot}$, so detected
galaxies are likely having biases bigger
than one.  Then, one can neglect the shot noise term $C_g^{\rm shot}$
with respect to $C_g$.  Taking the fact that $C^{\rm CMB}\gg C^{\rm
  ISW}$ (Fig. \ref{fig:ISW}),  the S/N of each $l$ is simplified to 
\be
\left(\frac{S}{N}\right)^2\simeq \frac{(2l+1)f_{\rm
    sky} r^2}{C_l^{\rm CMB}/C_l^{\rm ISW}}\ .
\ee
Here $r$ is  the cross correlation
coefficient between ISW and 
LSS. Since $r$ has very weak dependence on galaxy bias,
the estimation presented here is weakly model dependent. We 
disregard signals from $l< 20$, to reduce 
confusions of $\Lambda$CDM cosmology or dark energy models. For sparse
galaxy sampling which is sufficient for our purpose, SKA is able to
cover the whole sky. So we assume that 
$f_{\rm sky}=1$. 
The cumulative $\sum^{l_{\rm
    max}}_{20}(S/N)^2$ is 
shown in Fig.\ref{fig:ISW}.

The ISW  signal peaks at  $z\ga 3$ and increases toward high $l$. This
is hard to mimic by $\Lambda$CDM, dark energy or many forms of
modified gravity. (1) For $\Lambda$CDM or dark
energy models with $w\la -1$, at $z\ga 3$, the ISW effect effectively
vanishes.  Fig. \ref{fig:ISW} shows that $\Lambda$CDM can be
distinguished from the $\lambda_2=1000$ $f(R)$ gravity with $>2\sigma$
confidence  by the  
ISW-21cm emitting galaxy cross correlation.   
  (2) For 
dark energy models with $w\ga -1$, $A_{\rm ISW}$ does not
decrease as fast as that of $\Lambda$CDM. But the ISW signal (including
contributions from dark energy fluctuations) 
decreases toward high $l$ \cite{Hu04} and one does not expect a detectable ISW
effect. (3) DGP preserves the property of scale independent $D$
\cite{Lue04b,DGPperturbation}, so the ISW signal decreases toward high $l$,
like the dark energy case. Therefore we do not expect a detectable
signal at $l>20$ and $z>3$. 
 (4) For generalized $f\propto (\alpha 
R^2+\beta R_{ab}R^{ab}+\gamma R_{abcd}R^{abcd})^{-n}$ ($n>0$), the ISW
effect vanishes at high $z$ because the $f$ correction decreases much
faster than the exponential $f(R)$. So we expect that null detection
of ISW-LSS cross correlation at $l\geq 
20$ and $z\geq 3$ would constrain $\lambda_2$ to $\lambda_2>
1000$ at $>2\sigma$ confidence level. On the other hand, a detection
of such cross correlation would 
present as a severe challenge to general relativity.

{\it Discussion}.---
The scale dependence of $D$, as an unambiguous signature of modified
gravity, can in principle be measured from weak
gravitational lensing by the mean of lensing tomography. Since $\phi$
is no longer equal to $-\psi$, we provide the general form of the lensing
transformation matrix $A_{ij}$
\be
A_{ij}-\delta_{ij}=\int_0^{\chi_s}d\chi
(\phi-\psi)_{,ij}W(\chi,\chi_s)\ , 
\ee
where $W(\chi,\chi_s)=\chi(1-\chi/\chi_s)$ is the usual lensing
kernel. All basic lensing theorems remain unchanged. For example,
lensing shear field is still curl free ({\it if neglecting second
  order corrections such as Born correction}). For $f(R)$ gravity,
relation between the
lensing convergence $\kappa=1-(A_{11}+A_{22})/2$ and the matter
over-density resembles that of the general relativity, with
\be
\kappa=\frac{3}{2}H_0^2\Omega_0\int \delta
a^{-1}W(\chi,\chi_s)(1+f_R)^{-1}d\chi \ .
\ee
It is
interesting to see how well weak lensing alone can constrain modified
gravity. For the exponential $f(R)$, one complexity is that lensing
mainly probes LSS at $z\la 1$, 
where $Q$ is small and the deviation from a scale independent $D$ is
small, so the constraints may be weak. This can be significantly
improved by gravitational potential reconstructed from primary
CMB. Combining lensing and CMB measurements, it is very promising to
measure the 
evolution of the gravitational potential between $z=1100$ and $z\sim
0$ robustly. This will put strong constraints on the nature of
gravity. Unfortunately, due to singularities in the perturbation
equations, we are limited to scales $k\la 0.012h/$Mpc or $l\la 20$ at
$z\la 1$ (for $\lambda_2=1000$). Information contained in this region
is very limited and could be contaminated by other physics such as dark
energy fluctuations.   Solving the field equation crossing those
singularities  consistently 
is nontrivial. We leave this work for future study.

The $Q=1/2,2/3, 1$ singularities may be caused by awkward gauge choice, the
neglecting of time derivative terms with respect to corresponding
spatial derivative terms, or the
failure of the perturbation approach.  For example, for $Q=2/3$, the
only solution $\delta=0$ does not depend on initial conditions. This
could be caused by neglecting time derivative terms, which erases some
degrees of freedom. These issues require detailed study. But if these
singularities in LSS equations are physical, they can be applied to rule out many forms
of modified gravities as alternatives to dark energy or general
relativity. To produce a similar expansion history as those of dark
energy model, (1) $R$ should increase when $a$ decreases and  (2) 
$f(R(a=1))$ should be negative in order to mimic positive dark
energy. Furthermore, in order not to affect
inflation, BBN and primary CMB, $f(R(a\rightarrow 0))$ must be
sufficiently small. A sufficient (but not necessary) condition
satisfying the BBN constraint is that $f(R(a\rightarrow 0))\rightarrow 0$.  The   
exponential $f(R)$ and $1/R^{n}$ $f(R)$ all fall into this class. This
results in $f_R>0$ at least at some early epoch $a_+$. As we have seen
from previous discussions, $f_{RR}<0$ is a sufficient condition for the
existence of singularities. To  avoid singularities, $f_{RR}\geq 0$
must be satisfied  at all epochs. However, we will see that this requirement contradicts with
(1) and (2). $f_{RR}\geq 0$ results in $f_R(a<a_+)\geq f_R(a_+)>0$, because
 $R(a<a_+)>R(a_+)$. So, $f$ increases toward high redshift,
 crosses over zero at some epoch and then increases more quickly than
 $R$. Since  when $a\rightarrow 0$, $R\propto a^{-3}$, $f$ increases
 more quickly than $a^{-3}$ and thus more quickly than the matter density.
This violates condition (2). It could  have non-negligible effect on
BBN and contradicts our expectation.   
On the other hand, only for those $f(R)$ gravities in which
$f(R(a\rightarrow 0))$ does not vanish, singularities in LSS equations can be
avoided. A $\log R$ $f(R)$ gravity is such an example.  

To demonstrate the power of LSS to constrain gravity,  we adopt a
conservative requirement to 
avoid singularities at $k<k_s$.   At the limit that $\lambda_2\gg
1$,  $Q$ peaks at $a=(2\lambda_2/9\Omega_0)^{-1/3}$ and the peak
amplitude is $\simeq 12(1-\Omega_0)(2/9\Omega_0
 e)^{2/3}\lambda_2^{-4/3}(ck/H_0)^2$, where we show the speed of light
 $c$ explicitly. To avoid singularities at $k<k_s$, 
\be
\lambda_2\geq 2.5\times 10^5 \left(\frac{k_s}{h/{\rm Mpc}}\right)^{3/2}
\ee
should be satisfied.

{\it Acknowledgment}.--- I  thank Jochen Weller for providing the
Runge-Kutta code dverk.F and John Barrow,  Salvatore
Capozziello, Xuelei Chen, Tomi
Koivisto, Xinhe Meng, Sergei Odintsov and Roman Scoccimarro for helpful discussions. This 
work was supported in part by the One-Hundred-Talent Program of China
and by the DOE and the NASA grant NAG 5-10842 at Fermilab.


\begin{thebibliography}{}
\bibitem{Note_Neutrino} Neutrino free
  streaming causes $D$ to be scale dependent, but this effect is
  small. Fluctuations in dark energy cause $D$ to be scale dependent
  at $\sim$ horizon scale. But this effect vanishes toward smaller
  scales.
\bibitem{Solar} The condition $3\rho_{\rm solar}/\rho_c\ga 10^6$ is
  well  met for any experiments carried in the atmosphere of the
  Earth, since the atmosphere density is $\sim 10^{-3}$g$/$cm$^3$ and
  $\rho_c\simeq 10^{-29}$g$/$cm$^3$ ($h=0.7$). Though the matter density in the
  interplanetary space is not well measured, we have solid evidences
  that the condition can also be met for experiments carried out in
  the space. (1) The density of
  interplanetary dust (IPD) is $\rho_{\rm
  IPD}\ga 10^{-24}$g$/$cm$^3$ (\cite{Nieto05} and
  references therein). This gives $3\rho_{\rm solar}/\rho_c\ga 3\times
  10^6$. (2) The total mass of
  the Kuiper Belt objects between $40$ and $90$ AU  is
  $(0.6, 40)\times 10^{-5} M_{\oplus}$\cite{Teplitz98}. This converts
  to $3\rho_{\rm 
  solar}/\rho_c \ga 2\times 10^6$-$10^8$. (3) The
  anomalous Doppler frequency drift experienced by Pioneer 10 and
  11, if caused by drag force of interplanetary matter, requires the
  interplanetary matter density $\sim
  (5-30)\times 10^{-20}$g$/$cm$^3$ at 20-70 AU \cite{Nieto05}.  This
  converts to  $3\rho_{\rm solar}/\rho_c \ga 10^{10}$.
\end{thebibliography}
\end{document}